\documentstyle[epsfig,twocolumn,prb,aps]{revtex}

\begin{document}

\draft
\twocolumn[\hsize\textwidth\columnwidth\hsize\csname @twocolumnfalse\endcsname

\noindent

\title
{Anomalous Behavior of the Zero Field Susceptibility of
the Ising Model on the Cayley Tree} 

\author{Tatijana Sto\v si\' c$^{*}$, Borko D. Sto\v si\' c$^{\dag}$
and Ivon P. Fittipaldi$^{\ddag}$\\}
\bigskip
\bigskip

\address{$^{*}$Laboratory for Theoretical Physics, 
Institute for Nuclear Sciences, \\
Vin\v ca, P.O. Box 522, YU-11001 Belgrade, Yugoslavia}
\bigskip

\address{$^{\dag}$Departamento de F\' \i sica e Matem\' atica, 
Universidade Federal Rural de Pernambuco,\\
Rua Dom Manoel de Medeiros s/n, Dois Irm\~ aos,
52171-900 Recife-PE, Brasil
}
\bigskip

\address{$^{\ddag}$
Funda\c c\~ao de Amparo \' a Ci\^ encia e Tecnologia de Estado de Pernambuco - FACEPE\\
Rua Benfica 150, Madalena, 
50720-001 Recife-PE, Brasil
}

\date{August 21, 2002}


\maketitle
\begin{abstract}

It is found that the zero field susceptibility $\chi$ of the Ising model
on the Cayley tree exhibits unusually weak divergence at the
critical point $T_C$. The susceptibility amplitude is found to 
diverge at $T_C$ proportionally to the tree generation level $n$, while the 
behavior of $\chi$ is otherwise analytic in the vicinity of $T_C$, 
with the critical exponent $\gamma=0$.
\end{abstract}





\vskip1pc]


\narrowtext

It has been well established \cite{heimburg,matsuda,muller,morita}
that the Ising model on the Cayley tree exhibits no spontaneous order 
(zero magnetization in zero field), while the susceptibility diverges 
for temperatures lower than the critical value 
$T_C=2k_B^{-1}J/\ln [({\sqrt B}+1)/({\sqrt B}-1)]$, 
where $k_B$ is the Boltzmann constant, $J$ is the
nearest-neighbor interaction parameter, and $B$ is the tree branching
number (coordination number minus one). Above $T_C$, it was found through
several different {\it approximation schemes}
\cite{heimburg,matsuda,muller,morita} that the susceptibility asymptotically
behaves as
\begin{eqnarray}
\chi &&=
{\frac
{\beta \left( u+1 \right)^2}
{\left ( 1-B\,u^2\right )}
},
\label{intro1}
\end{eqnarray}
where $\beta={1}/{k_BT}$ and
$u= \tanh\left ({\beta J}\right)$.
Expanding the above expression around $t\equiv (T-T_C)/T_C$
one obtains
\begin{eqnarray}
\chi &&=
{\frac 
{\sqrt {B}+1} 
{2 J \sqrt {B} \left ( \sqrt {B}-1 \right ) }
} \; {t}^{-1}
+O\left (t^0\right )
\label{intro2}
\end{eqnarray}
corresponding to the ``classical" value of the critical exponent $\gamma=1$. The same 
value $\gamma=1$ was also found \cite{baxter,izmailian} for the Bethe lattice (interior 
of the Cayley tree), with the (higher) critical temperature 
$T_{BC}=2k_B^{-1}J/\ln \left[(B+1)/(B-1)\right]$.

On the other hand, in a recent work \cite{SSF} the present authors 
have established the {\it exact closed form 
expression} for the zero-field susceptibility $\chi\left ( n,T\right )$ of the Ising 
model on a Cayley tree of arbitrary generation $n$.
Our result \cite{SSF} confirmed previous findings \cite{heimburg,muller,morita}
concerning susceptibility high temperature behavior ($T > T_C$), 
but it also sheds new light on its singular behavior in the 
low temperature region $T\leq T_C$. 
In this work we perform a finite size scaling analysis of the
zero field susceptibility, paying special attention to its singular behavior
in the thermodynamic limit, at and below $T_C$.
In particular, it turns out 
that the singular behavior of $\chi$ at $T_C$, as displayed by the exact
formula, is quite different from that deduced from the approximate asymptotical
expression (\ref{intro1}) used in previous works \cite{heimburg,muller,morita}.
More precisely, we find that the terms of the order $\sim t^{-1}$ exactly cancel out, 
and it is shown that the susceptibility displays what may be termed a 
``divergent coincident singularity", in Fisher's classification \cite{fisher}, 
with critical exponent $\gamma=0$. 

For simplicity, we consider here only the Cayley tree with the branching number
$B=2$, the generalization to arbitrary $B$ being straightforward.
We further consider a single
$n$-generation branch of a Cayley tree, composed of two
$(n-1)$-generation branches connected to a single initial site, with
the Hamiltonian
\begin{equation}
{\cal H}=-J\sum_{\langle nn\rangle}S_iS_j-H\sum_{i}S_i,\label{one}
\end{equation}
where $H$ is the external magnetic field,
$S_i=\pm 1$ is the spin at site $i$, $\langle nn\rangle$ denotes
summation over the nearest-neighbor pairs, and $J$ denotes the 
coupling constant as before. The
$n$-generation branch consists of $N_n={2^{n+1}-1}$ spins, the
$0$-generation branch being a single spin.  The recursion relations for
the partition function of any two consecutive generation
branches were found by Eggarter \cite{egarter} to be
\begin{equation}
Z_{n+1}^{\pm}=y^{\pm 1}\left[ x^{\pm 1}Z_{n}^{+} 
+ x^{\mp 1}Z_{n}^{-}\right] ^2,\label{two}
\end{equation}
where $x\equiv \exp(\beta J)$, $y\equiv \exp(\beta H)$, and
$Z_{n}^{+}$ and $Z_{n}^{-}$ denote the branch partition functions
restricted by fixing the initial spin (connecting the two
($n-1$)-generation branches) 
into the $\{ +\}$ and $\{ -\}$ position, respectively. 

As the nonlinear coupled recursion relations (\ref{two}) can be iterated 
to yield a closed form 
expression \cite{egarter} only in the zero field case, several rather sophisticated
approximation schemes \cite{heimburg,matsuda,muller,morita} have been devised
to elaborate the field dependence of the partition function, and deduce (among other
quantities of interest) the limiting behavior of the zero field susceptibility 
{\it in the high temperature region}, as given by equation (\ref{intro1}).

On the other hand, only recently \cite{SSF} we have arrived at the exact analytical 
expression for the zero field susceptibility, by considering the recursion relations 
for the {\it field derivatives} of the partition function, which can be iterated in 
the limit $H\longrightarrow 0$ to yield corresponding closed form expressions. 
Zero field susceptibility of a Cayley tree of {\it arbitrary} generation $n$, as a 
function of temperature, was thus found \cite{SSF} to be

\begin{eqnarray}
\chi_n &&={\beta\over {2^{n+1}-1}}
\,\left [ {\frac {\left (u+1\right )^2\;{2}^{n+1}}{1-2\,{u}^{2}}}+
{\frac {u^2\; 2^{n+1}\; \left (2 u^2\right )^{n+1}}{(2\,u^2-1)(2\,u-1)^2}}
\right.\nonumber\\
&&\left.
+{\frac {2^{n+2}\; u^{n+2}}{(2\,u-1)^2}}
+{\frac {2\,{u}^{2}-1}{(2\,u-1)^2}}
\right ]. \label{eleven}
\end{eqnarray}

In the limit $n\longrightarrow \infty$ the first term ($h_n^{(1)}$) inside the 
square brackets on the right hand side of (\ref{eleven}) recovers formula 
(\ref{intro1}), and represents the dominant term for temperatures $T>T_C$. 
The second term ($h_n^{(2)}$) is dominant in the low temperature
region $T<T_C$, while the last two terms ($h_n^{(3)}$ and $h_n^{(4)}$) become 
negligible in the thermodynamic limit for all temperatures, and are therefore 
relevant only for finite size systems.

Up to this point everything matches the previous results and
conclusions of other authors \cite{heimburg,matsuda,muller,morita}, 
therefore, it came as somewhat of a surprise to find that upon expansion 
in power series  around 
$t=0^{\pm}$ the terms proportional to $t^{-1}$ 
{\it exactly cancel out, independent of the system generation level (therefore,
also in the thermodynamic limit), 
on both sides of the critical point}. It will be shown in the remainder of
this paper that this fact leads to quite different conclusions about
critical behavior of $\chi$, from those that follow from the approximate
expression (\ref{intro1}).

Expanding the four individual sequential terms, $h_n^{(1)}$, $h_n^{(2)}$, $h_n^{(3)}$ 
and $h_n^{(4)}$, inside the brackets
on the right hand side of (\ref{eleven}), in the vicinity
of $T_C$, and retaining only the terms of the order $t^{-1}$, it is found 

\begin{eqnarray}
h_n^{(1)}&&=
{\frac {\left( \sqrt{2}-1\right)^2\; 2^n}{\sqrt{2}\;{K_C}}}\;
t^{-1}+O(t^0) ,\cr
h_n^{(2)}&&=
-{\frac {\left( \sqrt{2}-1\right)^2\; 2^n}{\sqrt{2}\;{K_C}}}\;
t^{-1}+O(t^0) ,\cr
h_n^{(3)}&&=O(t^0) ,\cr
h_n^{(4)}&&=O(t^1) 
\label{twelve}
\end{eqnarray}
where $K_C\equiv J/k_B T_C=\ln{\left(1+\sqrt{2}\right)}\sim 0.881374$.
It is seen that the terms proportional to $t^{-1}$ 
exactly cancel out for arbitrary $n$, for both positive and negative $t$.
Retaining additional powers in $t$, for large $n$
one obtains

\begin{eqnarray}
\chi_n=
n\;K_C {\frac{\left(1+\sqrt{2}\right)^2}{2 J}}
\left[ 1-t\; n\; K_C\;
{\frac{1}{\sqrt{2}}}
+O(t^2)\right],
\label{thirteen}
\end{eqnarray}
which is valid on both sides of the critical point for $|t|<<1/n$ 
(since $t$ is an independent small parameter, this condition can
be fulfilled for arbitrarily large $n$). It is seen from (\ref{thirteen}) 
that in the thermodynamic limit $\chi_n$ diverges at $T_C$ proportionally 
to tree generation level $n$, rather then as $t^{-1}$ as follows from approximate 
formula (\ref{intro1}), {\it and is otherwise analytic} in $t$. 
It therefore follows that the susceptibility critical exponent is equal to $\gamma=0$,
while the {\it amplitude} diverges as $n\longrightarrow \infty$,
demonstrating behavior that may be termed \cite{fisher}
``divergent coincident singularity". It should be noted
that a Cayley tree of generation $n=77$ has a number of spins corresponding to
the Avogadro's number, while for $n\sim 270$ the number of spins corresponds to
the estimated number of hadrons in the observable Universe.
Therefore, the above susceptibility exhibits an extremely weak singularity.

To illustrate the behavior of susceptibility in the vicinity of $T_C$ in Fig.~1 we show
the function $\chi_n/n$, as given by formula (\ref{eleven}),
for several system sizes $n=64, 128, 256, 512$ and $1024$. 
All the shown curves $\chi_n/n$ intersect at 
$k_B T_C/J=1/\ln\left(1+\sqrt{2}\right)\sim 1.134593$,
exhibiting a finite value 
$\ln\left(1+\sqrt{2}\right)\;\left(1+\sqrt{2}\right)^2/2\sim 2.568511$.
Note that these correction terms turn already imperceptible on the scale of the graph.

\begin{figure}[htp]
{\hbox{\vbox{\psfig{figure=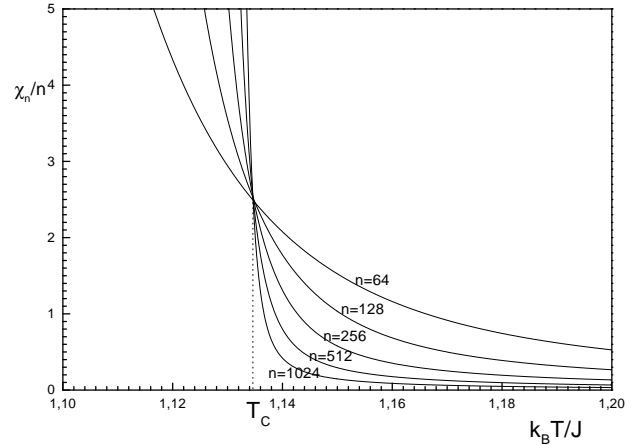,height=2.5in}}}}
\caption{
Zero field susceptibility as a function of temperature, 
calculated using formula (\ref{eleven}),
for several system sizes $n=32, 64, 126, 256$ and $1024$.
Crossing of the curves $\chi_n/n$ at $T_C$ demonstrates
the unusually weak singular behavior.
}
\label{fig1}
\end{figure}

In the temperature region below $T_C$ the second term ($h_n^{(2)}$)
inside the brackets on the right hand side of (\ref{eleven}) becomes dominant,
and for large $n$ susceptibility is given by

\begin{eqnarray}
\chi_n &&=\beta
{\frac {u^2\; \left (2 u^2\right )^{n+1}}{(2\,u^2-1)(2\,u-1)^2}}
. \label{fourteen}
\end{eqnarray}
It is seen that the susceptibility diverges as $\left(2u^2\right)^{n+1}$, 
the divergence becoming stronger as temperature is lowered. Scaling of susceptibility 
for different system sizes is obtained by taking the logarithm and then dividing by 
the generation level $n$. To demonstrate this scaling behavior, in Fig.~2 we show the 
function $\ln{\chi_n}/\left(n+1\right)$ for several system sizes, together with the 
corresponding limiting function $\ln\left ( 2 u^2\right )$. It is seen that the displayed 
finite size curves become hardly distinguishable from the limiting function for $0<T<T_C$.

\begin{figure}[htp]
{\hbox{\vbox{\psfig{figure=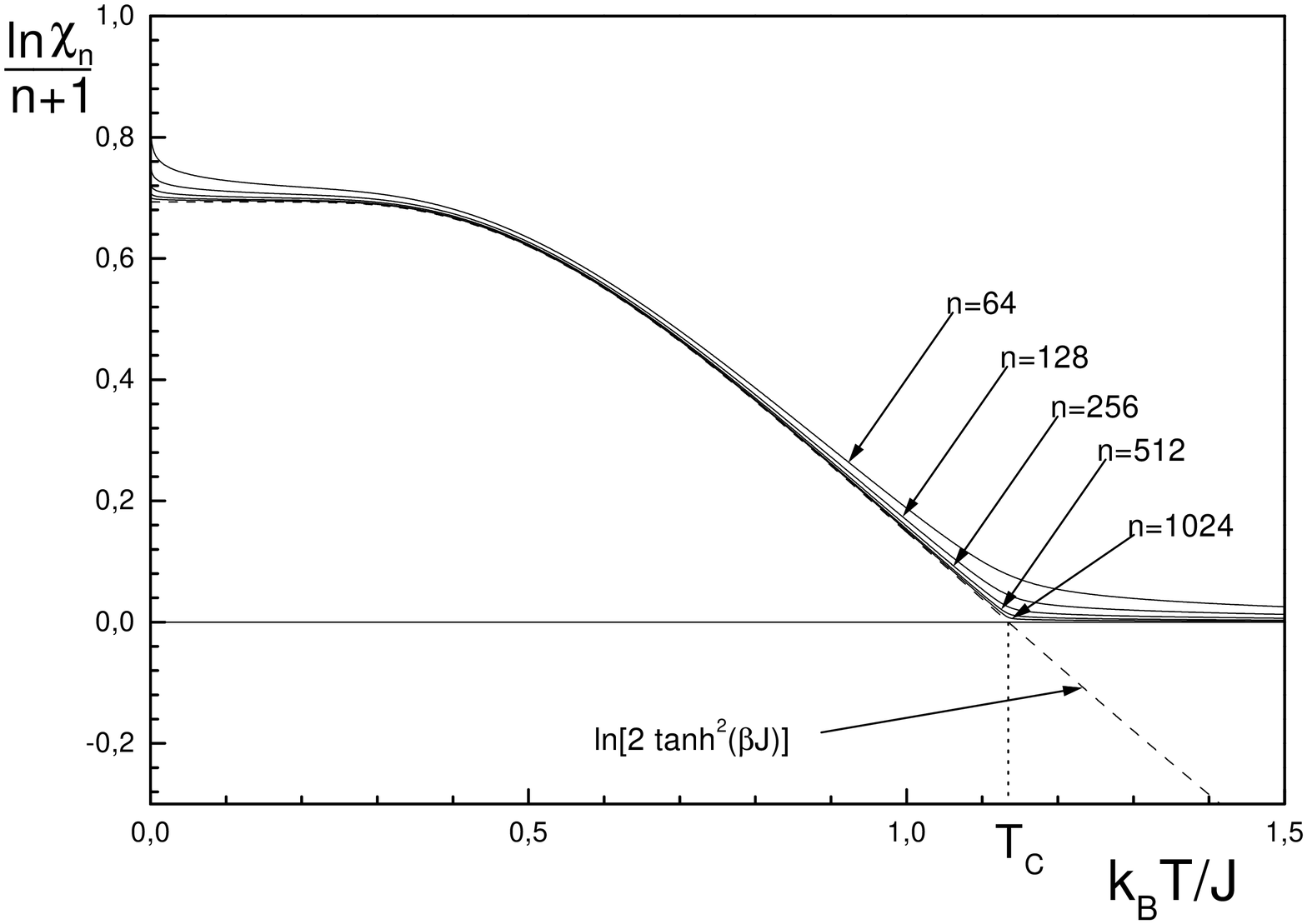,height=2.5in}}}}
\caption{
Finite size scaling of the zero field susceptibility in
the temperature region $0<T<T_C$.
The shown curves are calculated using formula (\ref{eleven}),
for several system sizes $n=32, 64, 126, 256$ and $1024$.
The dotted line represents the limiting curve 
$\ln\left[2\tanh^2(\beta J)\right]$.
}
\label{fig2}
\end{figure}

Finally, as $T$ approaches zero, the multiplicative
term $\beta$ takes over and susceptibility diverges trivially as $T^{-1}$,
for all generation levels.

In conclusion, the exotic structure of the Cayley tree, with its infinite dimension and
finite order of ramification, gives rise to rather unusual thermodynamic behavior. The overall
scaling behavior of susceptibility is governed by scaling of its amplitude, rather then
distance from the critical point. In particular, contrary to the conclusions drawn from the 
previous works\cite{heimburg,muller,morita} using the approximate formula (\ref{intro1}), 
at $T_C$ the susceptibility displays an extremely
weak singularity, with amplitude diverging proportionally to the tree generation level $n$
(as $n$ increases to infinity in the thermodynamic limit), 
with the critical exponent $\gamma=0$.

\section{Acknowledgements}
\bigskip
This work was supported in part by CNPq and FACEPE (Brazilian Agencies).

\end{document}